\documentclass[12pt]{article}

\usepackage{graphicx}
\usepackage{amsmath}
\usepackage{amssymb}

\newcommand{\la}{\lambda}
\newcommand{\K}{\mathrm{K}}
\newcommand{\E}{\mathrm{E}}
\newcommand{\sn}{\mathrm{sn}}

\newcommand{\prt}{\partial}

\begin{document}

\title{Wave breaking and the generation of undular bores in an integrable
shallow-water system}

\author{G.A. El$^{\dagger}$, R.H.J. Grimshaw$^{\ddagger}$, and A.M.~Kamchatnov$^{\S}$\\
$^{\dagger}$ School of Mathematical and Informational Sciences,\\
Coventry University, Priory Street, Coventry CV1 5FB, UK\\
$^{\ddagger}$ Department of Mathematical Sciences, Loughborough University,\\
Loughborough LE11 3T, UK \\
$^{\S}$ Institute of Spectroscopy, Russian Academy of Sciences,\\
Troitsk, Moscow Region,  142190 Russia}

\maketitle

\begin{abstract}
The generation of an undular bore in the vicinity of a wave-breaking
point is considered for the integrable Kaup-Boussinesq shallow water system.
In the framework of the Whitham modulation theory,
an analytic solution of the Gurevich-Pitaevskii type of  problem for
a generic ``cubic"  breaking regime is obtained
using a generalized hodograph transform, and a further reduction to a
linear Euler-Poisson equation.
The motion of the undular bore edges is investigated in detail.
\end{abstract}



\section{Introduction}

It is well known that the solutions  to an initial value problem for the
non-dispersive shallow water
equations may lead to wave-breaking after a finite time,
when the first spatial derivatives blow up.
After the wave breaking point,
a formal solution becomes multi-valued and loses its physical meaning.
The divergence of the spatial derivatives at the wave-breaking point
suggests that  dispersion effects described by terms with higher order
spatial derivatives must be taken into account.
Then these small dispersion effects lead to the
onset of oscillations in the vicinity of  the wave breaking point
followed by the development of an undular bore,
or in different terminology, a dissipationless shock wave.

This physical picture has been  put into mathematical form for waves
described by the Korteweg-de Vries (KdV) equation by
Gurevich and Pitaevskii (GP) in \cite{GP1} (see also Whitham in \cite{whitham2}).
In the GP formulation, the region of oscillations is presented as a
slowly modulated periodic wave solution of  the KdV equation.
The parameters of the wave change little on typical
wavelength/period scales
which permits one to apply  the Whitham modulation theory
(\cite{whitham1,whitham2}). The resulting Whitham equations describe
the slow evolution of the parameters in the undular bore.
In the original paper \cite{GP1},  two typical problems were
considered.
One problem is concerned with the description of decay  of an initial
discontinuity for the KdV equation and the exact $x/t$-similarity
solution of this problem was constructed.
Another problem corresponds
to the (universal) initial stage of development of a bore when the
solution of the dispersionless equation can be approximated
locally by a properly chosen cubic
curve. This problem was studied in \cite{GP1} numerically.
An exact analytic solution to this problem was later obtained by Pot\"emin
\cite{Potemin} using Krichever's
algebro-geometrical procedure for integration of the Whitham equations
\cite{Kr88}, \cite{DN}.
Later, Pot{\"e}min's solution was put into the general context of
Tsarev's generalized hodograph transform \cite{Tsarev1}
in \cite{KS, GKE, T}, where the hodograph equations were reduced to
the classical Euler-Poisson equation.

However, the KdV equation describes  unidirectional propagation of
nonlinear dispersive waves.
An integrable bi-directional analog of the KdV equation was
derived by Kaup in \cite{kaup} using
the Boussinesq approximation for shallow  water waves
\cite{whitham2}.  Like the KdV
equation, the Kaup-Boussinesq (KB) system is completely integrable and
therefore a powerful
inverse scattering transform method can be applied to its
investigation.
In particular, the multi-phase periodic solutions of the KB system were
found in \cite{MY}, the Whitham theory of modulations was applied in
\cite{EGP} to the problem of the decay of an initial discontinuity,
and a quasiclassical description of soliton trains arising from a
large initial pulse was developed in \cite{KKU}.

In this paper, we further extend the Gurevich-Pitaevskii theory to the
case of  bi-directional
shallow water equations using the KB system,
and construct an analytic solution to the
Whitham-KB equations for the
regime of generation of an undular bore in the vicinity of a breaking
point.  The obtained solution, along
with its own significance in the representation of undular bores,
will serve as an intermediate
asymptotic in a more general formulation we are presently
undertaking, in which small dissipation
is taken into account.

\section{Periodic waves in the Kaup-Boussinesq system}

In dimensionless units (see, e.g. \cite{EGP}) the KB system can be
written in the form
\begin{equation}
\begin{array}{l}
h_{t}+(hu)_{x}+ \frac{1}{4}u_{xxx}=0, \\
u_{t}+uu_{x}+h_{x}=0,
\end{array}
\label{eq1}
\end{equation}
where $h(x,t)$ denotes the height of the water surface above a horizontal
bottom and $u(x,t)$ is related to the horizontal velocity field (at the
leading order it is the depth-averaged horizontal field).

The KB system (\ref{eq1}) is completely integrable and can be
represented as the compatibility condition of two linear equations
\cite{kaup}
\begin{equation}
\psi _{xx}=\mathcal{A} \psi ,
\qquad
\psi _{t}=-\frac12\mathcal{B}_x\psi+\mathcal{B}\psi_x
\label{eq2}
\end{equation}
with
\begin{equation}\label{eq3}
    \mathcal{A}=\left(\la-\frac12 u\right)^2-h,\qquad
    \mathcal{B}=-\left(\la+\frac12 u\right).
\end{equation}
Thus, the inverse scattering transform method can be applied for its
investigation. In particular, the periodic solution of (\ref{eq1}) can
be obtained by the well-known finite-gap integration method (see, e.g.
\cite{kamch2000}) in the following way.
Let $\psi _{+}$ and $\psi _{-}$ be two basis solutions of the second
order spatial linear differential equation (\ref{eq2}). Then their product
\begin{equation}
g=\psi _{+}\psi _{-}  \label{eq4}
\end{equation}
satisfies the third order equation
\begin{equation}
g_{xxx}-2\mathcal{A}_xg-4\mathcal{A}g_x=0.
\label{eq5}
\end{equation}
Upon multiplication by $g$, this equation can be integrated once to
give
\begin{equation}
\frac12 gg_{xx}-\frac14 g_x^2-\mathcal{A}g^2=-P(\la)
\label{eq6}
\end{equation}
where the integration constant $P(\lambda )$ can only depend on
$\lambda $.
The time dependence of $g(x,t)$ is determined by the equation
\begin{equation}
g_{t}=\mathcal{B}g_x-\mathcal{B}_xg.
\label{eq7}
\end{equation}
This equation can readily be put in the form
\begin{equation}
\left( \frac{1}{g}\right)_{t}=\left( \frac{\mathcal{B}}{g}\right)_{x},
\label{eq8}
\end{equation}
which can in turn be considered as a generating function of an infinite
sequence of conservation laws.

The periodic solutions of the system (\ref{eq1}) are distinguished by
the condition that $P(\lambda )$ in (\ref{eq6}) be a polynomial in $\lambda $.
The one-phase periodic solution, which we are interested in,
corresponds to the fourth degree polynomial
\begin{equation}
P(\lambda )=\prod_{i=1}^{4}(\lambda -\lambda _{i})=\lambda
^{4}-s_{1}\lambda
^{3}+s_{2}\lambda ^{2}-s_{3}\lambda +s_{4}.
\label{eq9}
\end{equation}
Then we find from Eq.~(\ref{eq6}) that $g(x,t)$ is the first-degree
polynomial,
\begin{equation}
g(x,t)=\lambda -\mu (x,t),
\label{eq10}
\end{equation}
where $\mu (x,t)$ is connected with $u(x,t)$ and $h(x,t)$ by the
relations
\begin{equation}
u(x,t)=s_{1}-2\mu (x,t),\quad h(x,t)=\tfrac{1}{4}s_{1}^{2}-s_{2}-2\mu
^{2}+s_{1}\mu ,
\label{eq11}
\end{equation}
which in turn follow from a comparison of the coefficients of $\lambda ^{i}$ on both
sides of
Eq.~(\ref{eq6}). The spectral parameter $\lambda $ is arbitrary and on
substitution of $\lambda =\mu $ into Eq.~(\ref{eq6}) we obtain an
equation for $\mu $,
\[
\mu _{x}=2\sqrt{ P(\mu )},
\]
while a similar substitution into Eq.~(\ref{eq7}) gives
\[
\mu _{t}=-(\mu +\tfrac{1}{2}u)\mu _{x}=-\tfrac{1}{2}s_{1}\mu _{x}.
\]
Hence, $\mu (x,t)$ as well as $u(x,t)$ and $h(x,t)$ depend only on the
phase
\begin{equation}
\theta =x-\tfrac{1}{2}s_{1}t,
\label{eq12}
\end{equation}
\begin{equation}\label{eq12a}
\hbox{so that} \quad  V=\frac12 s_1=\frac12\sum_{i=1}^4\la_i
\end{equation}
is the phase velocity of the nonlinear wave,
and $\mu (\theta )$ is determined by the equation
\begin{equation}
\mu _{\theta }=2\sqrt{ P(\mu )}.
\label{eq13}
\end{equation}

For the fourth degree polynomial (\ref{eq9}) the solution of this
equation is
readily expressed in terms of elliptic functions. Let the zeros
$\lambda
_{i} $, $i=1,2,3,4,$ of the polynomial $P(\lambda )$ be real and
ordered
according to the rule
\begin{equation}
\lambda _{1}\leq\lambda _{2}\leq\lambda _{3}\leq\lambda _{4}.
\label{eq14}
\end{equation}
Then the real variable $\mu $ oscillates in the interval where the
expression under the square root in (\ref{eq13}) is positive,
\begin{equation}\label{eq15}
    \la_2\leq\mu\leq\la_3.
\end{equation}
Consequently the solution of Eq.~(\ref{eq13}) with the initial condition
$\mu(0)=\la_3$ is given by
\begin{equation}\label{eq16}
    \mu(\theta)=\frac{\la_3(\la_4-\la_2)-\la_4(\la_3-\la_2)\sn^2
    \left(\sqrt{(\la_4-\la_2)(\la_3-\la_1)}\,\theta,m\right)}
    {\la_4-\la_2-(\la_3-\la_2)\sn^2
    \left(\sqrt{(\la_4-\la_2)(\la_3-\la_1)}\,\theta, m \right)},
\end{equation}
where
\begin{equation}\label{eq17}
    m=\frac{(\la_3-\la_2)(\la_4-\la_1)}{(\la_4-\la_2)(\la_3-\la_1)}
\end{equation}
is the modulus of the elliptic functions. An equivalent solution
corresponding
to the initial condition $\mu(0)=\la_2$ is given by
\begin{equation}\label{eq18}
    \mu(\theta)=\frac{\la_2(\la_3-\la_1)-\la_1(\la_3-\la_2)\sn^2
    \left(\sqrt{(\la_4-\la_2)(\la_3-\la_1)}\,\theta,m\right)}
    {\la_3-\la_1-(\la_3-\la_2)\sn^2
    \left(\sqrt{(\la_4-\la_2)(\la_3-\la_1)}\,\theta,m\right)}.
\end{equation}
Substitution of (\ref{eq16}) or (\ref{eq18}) into (\ref{eq11}) gives
expressions for $u(\theta)$ and $h(\theta)$ in the periodic nonlinear
wave. Its wavelength is given by
\begin{equation}\label{eq19}
    L=\int_{\la_2}^{\la_3}\frac{d\mu}{\sqrt{P(\mu)}}=
    \frac{2\K(m)}{\sqrt{(\la_4-\la_2)(\la_3-\la_1)}},
\end{equation}
$\K(m)$ being the complete elliptic integral of the first kind.

The soliton limit $(m=1)$ is obtained either for $\la_1=\la_2$ or for
$\la_3=\la_4$.
For $\la_1=\la_2$ Eq.~(\ref{eq16}) gives
\begin{equation}\label{eq20}
    \mu(\theta)=\la_4-\frac{(\la_4-\la_1)(\la_4-\la_3)}
{\la_4-\la_1+(\la_3-\la_1)/\cosh^2[\sqrt{(\la_4-\la_1)(\la_3-\la_1)}\,\theta]},
\end{equation}
and  for $\la_3=\la_4$ Eq.~(\ref{eq18}) gives
\begin{equation}\label{eq21}
    \mu(\theta)=\la_1+\frac{(\la_2-\la_1)(\la_4-\la_1)}
{\la_2-\la_1+(\la_4-\la_2)/\cosh^2[\sqrt{(\la_4-\la_2)(\la_4-\la_1)}\,\theta]}.
\end{equation}
Their substitution into (\ref{eq11}) yields the soliton solution of the
KB system.

In the opposite limit $\la_2 = \la_3$ $(m=0)$ both expressions
(\ref{eq16}) and
(\ref{eq18}) reduce to
\begin{equation}\label{eq22}
    \mu=\la_2=\la_3\,.
\end{equation}
Thus the limit $\la_2 \to \la_3 $ yields sinusoidal waves.

\section{Whitham modulation equations for the Kaup-Boussinesq system}

The Whitham modulation equations describe the slow evolution of the
parameters $\la_i,\,i=1,2,3,4,$ of a modulated nonlinear wave.
They are
\begin{equation}\label{eq23}
    \frac{\prt\la_i}{\prt t}+v_i(\la)\frac{\prt\la_i}{\prt x}=0,
    \quad i=1,2,3,4,
\end{equation}
where the  Whitham velocities $v_i(\la)$ can be expressed in the form
\begin{equation}\label{eq24}
    v_i(\la)=\left(1-\frac{L}{\prt_iL}\prt_i\right)V,\quad
    \prt_i\equiv\frac{\prt}{\prt\la_i},\quad i=1,2,3,4,
\end{equation}
where the phase velocity $V$ and the wavelength $L$ are given
correspondingly by (\ref{eq12a}) and (\ref{eq19}). A simple
calculation yields the explicit expressions \cite{EGP},
\begin{equation}\label{eq25}
    \begin{array}{l}\displaystyle{
    v_1=\frac12\sum\la_i-\frac{(\la_4-\la_1)(\la_2-\la_1)\K(m)}
    {(\la_2-\la_1)\K(m)+(\la_4-\la_2)\E(m)},}\\\displaystyle{
    v_2=\frac12\sum\la_i-\frac{(\la_3-\la_1)(\la_2-\la_1)\K(m)}
    {(\la_2-\la_1)\K(m)-(\la_3-\la_1)\E(m)},}\\\displaystyle{
    v_3=\frac12\sum\la_i+\frac{(\la_4-\la_3)(\la_3-\la_2)\K(m)}
    {(\la_4-\la_3)\K(m)-(\la_4-\la_2)\E(m)},}\\\displaystyle{
    v_4=\frac12\sum\la_i+\frac{(\la_4-\la_3)(\la_4-\la_1)\K(m)}
    {(\la_4-\la_3)\K(m)+(\la_3-\la_1)\E(m)},}
    \end{array}
\end{equation}
where $\K(m)$ and $\E(m)$ are complete elliptic integrals of
the first and second kind, respectively.

In the limit $\la_1=\la_2$ $(m=1)$ the Whitham velocities reduce to
\begin{equation}\label{eq26}
    v_1=v_2=\frac12\sum\la_i,\quad v_3=\frac12(3\la_3+\la_4),
    \quad v_4=\frac12(\la_3+3\la_4);
\end{equation}
in the limit $\la_3=\la_4$ $(m=1)$ they reduce to
\begin{equation}\label{eq27}
    v_1=\frac12(3\la_1+\la_2),\quad v_2=\frac12(\la_1+3\la_2),
    \quad v_3=v_4=\frac12\sum\la_i;
\end{equation}
and in the limit $\la_2=\la_3$ $(m=0)$ they reduce to
\begin{equation}\label{eq28}
    v_1=\frac12(3\la_1+\la_4),\quad v_2=v_3=\frac12\sum\la_i+
    \frac{2(\la_2-\la_1)(\la_4-\la_2)}{\la_4+\la_1-2\la_2},
    \quad v_4=\frac12(\la_1+3\la_4).
\end{equation}

Next we shall apply the Whitham theory to the description of the undular
bore forming in the vicinity of a
wave-breaking singularity.

\section{The Gurevich-Pitaevskii problem for the KB system}

\subsection{Wave breaking in the dispersionless limit}

In the dispersionless limit, the KB system (\ref{eq1}) reduces to
well-known
shallow water equations
\begin{equation}\label{eq30}
    h_t+(hu)_x=0,\qquad u_t+uu_x+h_x=0,
\end{equation}
which can be transformed to the diagonal form
\begin{equation}\label{eq31}
    \frac{\prt\la_+}{\prt t}+\frac12(3\la_++\la_-)\frac{\prt\la_+}{\prt
x}=0,
    \quad \frac{\prt\la_-}{\prt
t}+\frac12(\la_++3\la_-)\frac{\prt\la_-}{\prt x}=0,
\end{equation}
\begin{equation}\label{eq32}
\hbox{where} \quad  \la_{\pm}=\frac{u}2\pm\sqrt{h}
\end{equation}
are the Riemann invariants of Eqs.~(\ref{eq30}).

Initial data are given by two functions $\la_+(x,0)$ and $\la_-(x,0)$
determined by the initial distributions $h_0(x)$ and $u_0(x)$. The
system (\ref{eq31}) has two families of characteristics
in the $(x,t)$ plane along which one of two Riemann invariants
(either $\la_+$ or $\la_-$) is constant. The wave-breaking point
corresponds to the moment when characteristics of one of the
families begin to intersect, so that the corresponding
Riemann invariant becomes a three-valued function in the physical
plane.
Let such an intersection occur for the characteristics transferring the
values of $\la_+$. Then at the wave-breaking point the profile of
$\la_+$ as a function of $x$ has a vertical tangent line and, hence,
in vicinity of this point it varies very fast, whereas the second
Riemann invariant varies with $x$ more slowly and may be
considered here as a constant parameter:
\begin{equation}
\label{eq33}
\la_-=\la_0={\rm const}.
\end{equation}
Thus, in the vicinity of the breaking point we are dealing with a  simple
wave. The second equation in (\ref{eq31}) is identically  satisfied by
Eq.~(\ref{eq33}).
The first equation in (\ref{eq31}) then has the well-known solution
\begin{equation}
x-\frac12(3\la_++\la_-)t=f(\la_+),
\end{equation}
where $f(\la_+)$ is an inverse function to an initial profile
$\la_+(x,0)$.
At the wave-breaking time, normalized here to be $t=0$, the function
$x=f(\la_+)$ must have an inflexion point with a vertical tangent
line.  Then in the vicinity of this point $f(\la_+)$ can be approximated by a
cubic function,
\begin{equation}
\label{eq35}
x-\frac12(3\la_++\la_-)t=-C(\la_+- \overline{\la}_1)^3.
\end{equation}
The KB system (\ref{eq1}) is invariant with respect to Galilean
transformation
\begin{equation}\label{eq36}
    x'=x-u_0t,\quad t'=t,\quad h=h',\quad u=u'+u_0,\quad
\la_{\pm}=\la_{\pm}'+\frac{u_0}2,
\end{equation}
and scaling transformation
\begin{equation}\label{eq37}
    x=ax',\quad t=a^2t',\quad h=h'/a^2,\quad u=u'/a,\quad \la_{\pm}=\la_{\pm}'/a.
\end{equation}
With the aid of these transformations Eq.~(\ref{eq35}) can be cast into
the form
\begin{equation}\label{eq38}
    x-\frac12(3\la_++\la_-)t=-\la_+^3,\quad \la_-=\lambda_0 \, .
\end{equation}
where we  have omitted the  prime superscripts for notational convenience.
It corresponds to the wave breaking picture shown in Fig.~1.
\begin{figure}[ht]
\centerline{\includegraphics[width=8cm,height=5cm,clip]{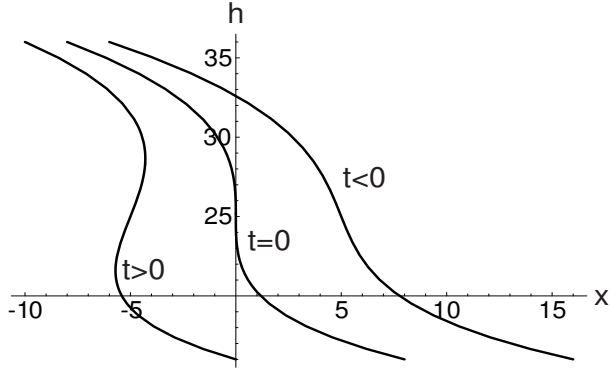}}
\vspace{0.3 true cm}
\caption{Wave breaking of the water elevation in the dispersionless
limit;
$\la_-$ is taken equal to -10.
}
\label{figone}
\end{figure}

The actual solution of the KB system now consists of two parts.
Following Gurevich and Pitaevskii \cite{GP1}, we suppose that
the region of oscillations can be approximated by a modulated
periodic solution of the KB system. Its evolution is determined by
the Whitham equations (\ref{eq23}) and we have to find that
solution which matches the solution (\ref{eq38}) at the end points
of the oscillatory region. One may say that this oscillatory region
 (the undular bore) ``replaces'' a
non-physical multi-valued region of the solution (\ref{eq38}). One
should emphasize, however, that
the boundaries of the undular bore {\it do not coincide} with the
boundaries of formal multi-valued solution.
Outside these boundaries, the solution approaches  the dispersionless
solution (\ref{eq38}).

\subsection{Undular bore solution}

We look for the solution of the Whitham equations (\ref{eq23})
in the form
\begin{equation}\label{eq39}
    x-v_i(\la)t=w_i(\la),\quad i=1,2,3,4,
\end{equation}
where $v_i$ are the Whitham velocities (\ref{eq25}). Since we consider
the breaking of the Riemann invariant $\la_+$ and $\la_-<\la_+$, we
take
\begin{equation}\label{eq40}
    \la_1=\la_-=\la_0=\mathrm{const}.
\end{equation}
Then the limiting formulas (\ref{eq26})-(\ref{eq28}) show that if we
find  $w_i(\la)$ such that
\begin{equation}\label{eq41}
    \begin{split}
    w_4=-\la_4^3\quad \mathrm{at}\quad \la_2=\la_3\quad (m=0),\\
    w_2=-\la_2^3\quad \mathrm{at}\quad \la_4=\la_3\quad (m=1),
    \end{split}
\end{equation}
then Eqs.~(\ref{eq39}) determine the Riemann invariants in such a way
that $\la_4=\la_+$ at the trailing edge $x^-(t)$ where $m=0$,
$\la_2=\la_+$
at the leading edge $x^+(t)$ where $m=1$, and $\la_1=\la_-=\la_0$
everywhere.
Thus, the plots of Riemann invariants $\la_2,\la_3,\la_4$ as functions
of $x$
are joined into continuous curve whose upper and lower branches match
with
the solution (\ref{eq38}) of the dispersionless equations (see Fig.~2).
In the region $x^-(t)<x<x^+(t)$ there are four Riemann invariants which
determine the modulated periodic solution representing the undular
bore.
At  its trailing edge $x\to x^-(t)$ the amplitude of oscillations
vanishes and at the leading edge $x\to x^+(t)$ the periodic solution
transforms
into a soliton train.
\begin{figure}[ht]
\centerline{\includegraphics[width=8cm,height=5cm,clip]{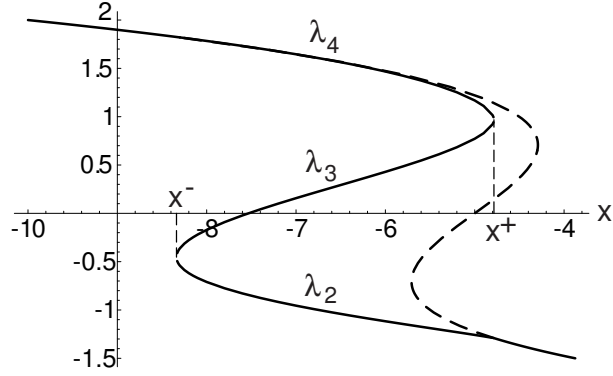}}
\vspace{0.3 true cm}
\caption{Dependence of Riemann invariants $\la_2,\la_3,\la_4$ on $x$
at fixed value of time $t=1$ and with $\la_1=-10$. The dashed line
shows the corresponding dependence of $\la_+$ for the solution of
the KB equations in the dispersionless limit.
}
\label{figtwo}
\end{figure}

According to the generalized hodograph method \cite{Tsarev1, Tsarev2},
Eqs.~(\ref{eq39})
satisfy the Whitham equations (\ref{eq23}) provided $w_i(\la)$ are the
velocities of the flows
\begin{equation}\label{eq42}
    \frac{\prt\la_i}{\prt \tau}+w_i(\la)\frac{\prt\la_i}{\prt x}=0,
    \quad i=1,2,3,4,
\end{equation}
commuting with (\ref{eq23}), i.e $\partial_{t \tau}\la_i= \partial_{
\tau t}\la_i$. If we represent $w_i(\la)$ in the
form analogous to Eqs.~(\ref{eq24}),
\begin{equation}
\label{eq43}
w_i(\la)=\left(1-\frac{L}{\partial_iL}\partial_i\right)W,\quad
 i=1,2,3,4,
\end{equation}
then the condition of commutativity of the flows (\ref{eq23}) and
(\ref{eq43}) reduces to the system of Euler-Poisson equations, exactly
as happens in the KdV  \cite{KS, GKE, T}
and NLS \cite{EK95} cases,
\begin{equation}
\label{eq44}
\partial_i\partial_jW-\frac1{2(\la_i-\la_j)}(\partial_iW-\partial_jW)=0,\quad
i\neq j.
\end{equation}
It is easy to check that this equation has a particular solution
$W={\mathrm{const}}/\sqrt{P(\la)},$ $P(\la)=\prod(\la-\la_i),$ which is
sufficient for our purpose. We choose the normalization factor so that
the coefficient before $\la^{-1}$ in the series expansion of $W$ in
powers of $\la^{-1}$ be equal to the phase velocity of the periodic
wave $s_1/2=V$.  Thus, we
obtain the sequence of $W^{(k)}$ defined by the generating
function
\begin{equation}
\label{eq45}
W=\frac{\la^2}{\sqrt{P(\la)}}=\sum\frac{W^{(k)}}{\la^k}=
1+\tfrac12{s_1}\cdot\frac1{\la}+
\left(\tfrac38s_1^2-\tfrac12s_2\right)\cdot\frac1{\la^2}
+\left(\tfrac5{16}s_1^3-\tfrac34s_1s_2+\tfrac12s_3\right)
\cdot\frac1{\la^3}+\ldots.
\end{equation}
Next a sequence of velocities of the commuting flows is given by
\begin{equation}
\label{eq46}
w_i^{(k)}(\la)=\left(1-\frac{L}{\partial_iL}\partial_i\right)W^{(k)},\quad
 i=1,2,3,4,
\end{equation}
where $w_i^{(1)}=v_i$ coincide with the Riemann velocities
(\ref{eq25}).
It is not difficult to find the limiting formulas analogous to
(\ref{eq26})-(\ref{eq28}). In particular, we get at $\la_3=\la_4$
$(m=1)$
\begin{equation}\label{eq47}
   \begin{split}
    &\left.w_1^{(1)}\right|_{\la_3=\la_4}=\left.w_2^{(1)}\right|_{\la_3=\la_4}
    =\frac12(\la_1+3\la_2),  \\
    &\left.w_1^{(2)}\right|_{\la_3=\la_4}=\left.w_2^{(2)}\right|_{\la_3=\la_4}
    =\frac38(\la_1^2+2\la_1\la_2+5\la_2^2),  \\
    &\left.w_1^{(3)}\right|_{\la_3=\la_4}=\left.w_2^{(3)}\right|_{\la_3=\la_4}
    =\frac1{16}(5\la_1^3+9\la_1^2\la_2+15\la_1\la_2^3+35\la_2^3).
    \end{split}
\end{equation}
Now we take such the linear combination
$$
w_2=a_0+a_1w_2^{(1)}+a_2w_2^{(2)}+a_3w_2^{(3)}
$$
so that $w_2$ satisfies the condition (\ref{eq41}). The coefficients
$a_1,a_2,a_3,a_4$
depend on the constant Riemann invariant $\la_1=\la_-=\la_0$ and their
values found
in this way yield the required solution of the Whitham equations:
\begin{equation}
\label{eq48}
\begin{split}
x-v_i(\la)t&=-\tfrac{16}{35}w_i^{(3)}(\la)+\tfrac8{35}{\la_0}w_i^{(2)}+
\tfrac2{35}{\la_0}^2v_i(\la)+\tfrac1{35}{\la_0}^3,\quad i=2,3,4;\\
\la_1&={\la_0}={\rm const}.
\end{split}
\end{equation}
These formulas define  $\la_2,\la_3,\la_4$ implicitly  as functions of
$x$ and $t$
and give the solution of the Gurevich-Pitaevskii problem for the
KB-Whitham system.
It is interesting to note that, unlike the KdV case, this solution
is not scale-invariant (we recall that
in the counterpart KdV solution $\la_i=t^{-1/2}l_i(x/t^{3/2}),
i=1,2,3$).  This  happens due to the
presence of the fourth Riemann invariant $\lambda_1$ in the Whitham
equations which is constant
($\lambda_0$) for the obtained solution and cannot be eliminated by
simple Galilean transform.
As a result, the solution (\ref{eq48}) does not possess the scaling
invariance required for a
generalized similarity behaviour  of $\lambda_j\, , \  j=2,3,4$. The
only family of admissible such similarity ($x/t$) solutions is
realized in the simplest case of the decay of an initial discontinuity
studied in \cite{EGP}.

\subsection{Laws of motion at the trailing and leading edges of the
oscillatory region}

Let us find the laws of motion at the leading and trailing edges of the
undular bore. First we consider the leading edge
$\la_3=\la_4$
$(m=1)$, and define the small deviations $\la_3'$ and $\la_4'$
from the value $\la_3=\la_4=\la_4^+$:
\begin{equation}\label{eq49a}
    \la_3=\la_4^++\la_3',\qquad \la_4=\la_4^++\la_4' \,.
\end{equation}
Then we seek the asymptotic expressions for formulas (\ref{eq48}) with
$i=3,4$
for small $|\la_3'|,|\la_4'|$:
\begin{equation}\label{eq49}
    x^++x'-(v_3^++v_3')t=w_3^++w_3',\quad
    x^++x'-(v_4^++v_4')t=w_4^++w_4',
\end{equation}
where $x'$ denotes the space coordinate reckoned from its limiting
value
$x^+$ and
\begin{equation}\label{eq50}
    v_3^+=v_4^+=\tfrac12(\la_1+\la_2+2\la_4),
\end{equation}
\begin{equation}\label{eq51}
v_3'=-v_4'=-\frac12\left\{\la_3'\ln\left[\frac{-(\la_2-\la_1)\la_3'}
    {(\la_4-\la_1)(\la_4-\la_2)}\right]-
    \la_4'\ln\left[\frac{(\la_2-\la_1)\la_4'}
    {(\la_4-\la_1)(\la_4-\la_2)}\right]\right\},
\end{equation}
\begin{equation}\label{eq52}
w_3^+=w_4^+=-\frac1{35}(5\la_2^3+6\la_2^2\la_4+8\la_2\la_4^2+16\la_4^2),
\end{equation}
\begin{equation}\label{eq53}
\begin{split}
    w_3'=-w_4'=&\frac1{35}(3\la_2^2+8\la_2\la_4+24\la_4^2)\times\\
    &\left\{\la_3'\ln\left[\frac{-(\la_2-\la_1)\la_3'}
    {(\la_4-\la_1)(\la_4-\la_2)}\right]-
    \la_4'\ln\left[\frac{(\la_2-\la_1)\la_4'}
    {(\la_4-\la_1)(\la_4-\la_2)}\right]\right\},
    \end{split}
\end{equation}
while  $\la_1,\la_2,\la_4$ denote here their limiting values
$\la_1^+,\la_2^+,\la_4^+$, correspondingly. Then subtraction of
one equation (\ref{eq49}) from the other yields at once the
relationship
\begin{equation}\label{eq54}
    t=\frac2{35}(3\la_2^2+8\la_2\la_4+24\la_4^2).
\end{equation}
On the other hand, the limiting formulas
$$
x^+-v_2^+t=w_2^+,\quad x^+-v_3^+t=w_3^+,
$$
with
\begin{equation}\label{eq55}
    v_2^+=\tfrac12(\la_1+3\la_2),\qquad w_2^+=-\la_2^3
\end{equation}
and $v_3^+$, $w_3^+$ given by (\ref{eq50}) and (\ref{eq52}) give after
subtraction the relationship
\begin{equation}\label{eq56}
    t=\frac2{35}(15\la_2^2+12\la_2\la_4+8\la_4^2).
\end{equation}
Then equating of the right hand sides of (\ref{eq54}) and (\ref{eq56})
shows that at the leading edge we have
\begin{equation}\label{eq57}
    \la_4^+=-\frac34\la_2^+.
\end{equation}
Substitution of this relation into (\ref{eq54}) or (\ref{eq56}) gives
the
dependence of $\la_2^+$ on $t$:
\begin{equation}\label{eq58}
    \la_2^+=-\left(\frac{5t}3\right)^{1/2}.
\end{equation}
At last, the formula $x^+-v_2^+t=-(\la_2^+)^3$ yields the law of motion
for the leading edge:
\begin{equation}\label{eq59}
    x^+(t)=\frac12\la_0t+\frac16\sqrt{\frac53}\,t^{3/2}.
\end{equation}

In a similar way we can consider  the trailing edge
$\la_3=\la_2=\la_2^-$
$(m=0)$ where we  define
\begin{equation}\label{eq60}
    \la_2=\la_2^-+\la_2',\qquad \la_3=\la_2^-+\la_3'
\end{equation}
and Eqs.~(\ref{eq48}) with $i=2,3$ reduce to
\begin{equation}\label{eq61}
    x^-+x'-(v_2^++v_2')t=w_2^++w_2',\quad
    x^-+x'-(v_3^++v_3')t=w_3^++w_3',
\end{equation}
where
\begin{equation}\label{eq62}
v_2^-=v_3^-=\frac{\la_1^2+4\la_1\la_2-8\la_2^2-2\la_1\la_4+4\la_2\la_4+\la_4^2}
    {2(\la_1-2\la_2+\la_4)},
\end{equation}
\begin{equation}\label{eq63}
    v_2'=-v_3'=-\frac{\la_2'+3\la_3'}{2(\la_1-2\la_2+\la_4)}
    (3\la_1^2-8\la_1\la_2+8\la_2^2+2\la_1\la_4-8\la_2\la_4+3\la_4^2),
\end{equation}
\begin{equation}\label{eq64}
    \begin{split}
w_2^-=w_3^-=&[128\la_2^4-64\la_2^3\la_4-16\la_2^2\la_4^2-8\la_2\la_4^3-5\la_4^4\\
&-7\la_1(16\la_2^3-8\la_2^2\la_4-2\la_2\la_4^2-\la_4^3)]/(35(\la_1-2\la_2+\la_4))
    \end{split}
\end{equation}
\begin{equation}\label{eq65}
    \begin{split}
w_2'=-w_3'=\frac{\la_2'+3\la_3'}{70(\la_1-2\la_2+\la_4)^2}[&-384\la_2^4
    +384\la_2^3\la_4-80\la_2^2\la_4^2-16\la_2\la_4^3-9\la_4^4\\
    &+7\la_1^2(-24\la_2^2+8\la_2\la_4+\la_4^2)].
    \end{split}
\end{equation}
Then subtraction of one equation (\ref{eq61}) from the other gives
\begin{equation}\label{eq66}
\begin{split}
t=\tfrac2{35}[&-384\la_2^4+384\la_2^3\la_4-80\la_2^2\la_4^2-16\la_2\la_4^3-9\la_4^4\\
&+7\la_1^2(-24\la_2^2+8\la_2\la_4+\la_4^2)]/(3\la_1^2-8\la_1\la_2+8\la_2^2
    +2\la_1\la_4-8\la_2\la_4+3\la_4^2).
    \end{split}
\end{equation}
Now the limiting formulas
\begin{equation}\label{eq67}
    x^--\frac12(\la_1+3\la_4)t=-\la_4^3,\qquad x^--v_3^-t=-w_3^-
\end{equation}
give
\begin{equation}\label{eq68}
t=\frac2{35}\frac{(8\la_2-7\la_1)(8\la_2^2+4\la_2\la_4+3\la_4^2)-15\la_4^3}
{4\la_2-3\la_1-\la_4}.
\end{equation}
Equating the right hand sides of (\ref{eq66}) and (\ref{eq68}), we find
the
relationship between the values of $\la_2^-$ and $\la_4^-$ at the
trailing edge:
\begin{equation}\label{eq69}
    \begin{split}
    21\la_1^2(\la_4+4\la_2)-10\la_1(20\la_2^2+2\la_2\la_4-\la_4^2)
    +16(8\la_2^3-\la_2^2\la_4-\la_2\la_4^2)+9\la_4^3=0.
    \end{split}
\end{equation}
Given $t$ and $\la_1=\la_0=\mathrm{const}$, we find $\la_2=\la_2^-$ and
$\la_4=\la_4^-$ from (\ref{eq68}) and (\ref{eq69}) and then the law of
motion of the trailing edge follows from
\begin{equation}\label{eq70}
    x^-=\frac12(\la_1+3\la_4)t-\la_4^3.
\end{equation}

\begin{figure}[ht]
\centerline{\includegraphics[width=8cm,height=5cm,clip]{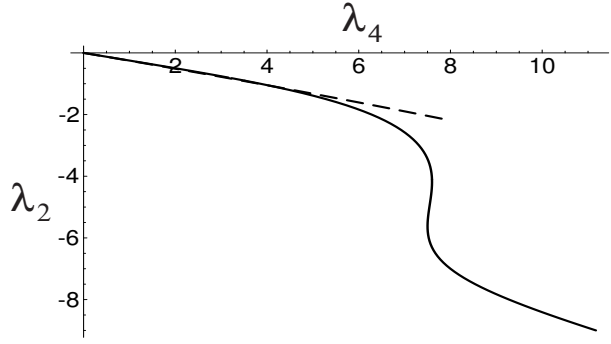}}
\vspace{0.3 true cm}
\caption{Riemann invariant $\la_2$ as a function of $\la_4$ defined
implicitly by Eq.~(\ref{eq69}). The plot corresponds to a fixed
value of $\la_1=-10$. Dashed line shows the dependence according to
asymptotic formula (\ref{eq73}).
}
\label{figthree-a}
\end{figure}

It is worth noticing that Eq.~(\ref{eq57}) coincides with the
corresponding
relation between the Riemann invariants at the leading edge in the
solution
of the Gurevich-Pitaevskii problem in the KdV equation case (see, e.g.
\cite{kamch2000}). A similar relation $\la_2^-=-\la_4^-/4$ between the
Riemann
invariants of the KdV theory follows from (\ref{eq69}) in the limit
$|\la_1|=|\la_0|\to\infty$. In the next approximation we obtain
\begin{equation}\label{eq71}
    \la_2=\la_3\cong-\frac14\la_4-\frac5{168}\frac{\la_4^2}{\la_1},
\end{equation}
and similar expansion of Eq.~(\ref{eq68}) in powers of $1/\la_1$,
\begin{equation}\label{eq72}
    t\cong
\frac13\la_4^2\left(1+\frac{10}{21}\frac{\la_4}{\la_1}\right),
\end{equation}
yields with the same accuracy
\begin{equation}\label{eq73}
    \la_4\cong\sqrt{3t}-\frac57\frac{t}{\la_1},\qquad
\sqrt{3t}\ll|\la_1|.
\end{equation}
Dependence of $\la_2$ on $\la_4$ given by Eq.~(\ref{eq69}) with fixed
value of $\la_1$ is illustrated in Fig.~\ref{figthree-a}.

\begin{figure}[ht]
\centerline{\includegraphics[width=8cm,height=5cm,clip]{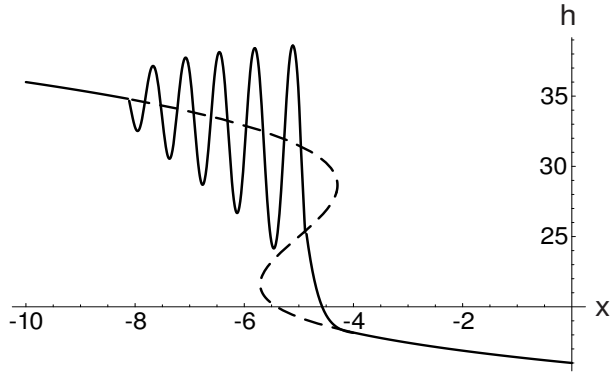}}
\vspace{0.3 true cm}
\caption{Undular bore for the KB equations. The plot
corresponds to the time $t=1$ and $\la_1=-10$. Dashed line shows the
solution in the dispersionless limit.
}
\label{figthree}
\end{figure}

Substitution of Eq.~(\ref{eq72}) into Eq.~(\ref{eq70}) yields
an approximate expression for the law of motion of the trailing edge;
\begin{equation}\label{eq74}
    x^-\cong\frac12\la_0t-\frac{3\sqrt{3}}2 t^{3/2}+\frac{75}{14}
    \frac{t^2}{\la_0},\qquad \sqrt{3t}\ll|\la_0|.
\end{equation}
Thus, the analytic formulas (\ref{eq48}) for the solution of the
Whitham
equations allowed us to find the main characteristics of the
dissipationless shock. With the use of Eqs.~(\ref{eq48}) we can find
$\la_2,\la_3,\la_4$ as functions of $x$ at given $t$, and their
substitution into (\ref{eq16}) and (\ref{eq11}) yields the profiles
of $u(x)$ and $h(x)$ in the undular bore. An example of such
a profile of the water elevation is shown in Fig.~\ref{figthree}.
As we see, the non-physical solution obtained in the dispersionless
limit is replaced by an undular bore. Its end points move
according to the laws found above, so that the oscillatory region
expands with time and its width grows mainly as $t^{3/2}$.
Small amplitude oscillations are generated at the trailing edge
(actually these oscillations represent gravity waves propagating
into an undisturbed smooth region), and they transform gradually
into solitons at the leading edge.

\section{Conclusion}

In this article, the wave breaking (Gurevich-Pitaevskii) problem is
solved for the shallow water waves described by the Kaup-Boussinesq
system.
This theory generalizes the KdV model to the case of a
bi-directional
wave propagation model, and so allows for the effects of wave
interaction.
As a result, the dispersionless
theory includes two Riemann invariants and the breaking of the wave
means breaking of one of these two invariants. Thus, the wave
breaking picture depends on the value of an additional  parameter:
the value of the  non-breaking  Riemann invariant. Correspondingly, the
Whitham theory for the
KB system contains four Riemann invariants (instead of three in the KdV
case) and,
as a result, the solution of the
Whitham equations for the Gurevich-Pitaevskii problem of the wave front
breaking is parameterized by a constant value $\la_0$.

The analytic solution of the
Whitham equations obtained here allows one to investigate the form of an undular bore
as a function of the parameter $\la_0$. In particular, the velocity of the
trailing edge increases with decrease of $|\la_0|$ and in the
limit $|\la_0|\to\infty$ the present theory reduces to known
results of the KdV model.

Our obtained solution can also be viewed as an
intermediate asymptotic in a more general
problem of the description of frictional shallow water undular bores where
small dissipation is taken into account (see \cite{GP2,AKN,MG} for the
results relevant to the KdV equation with weak dissipation).  This
problem will be the subject of a separate study.

\subsection*{Acknowledgements}
This work was completed during stay of A.M.K. at Department of Mathematical
Sciences, Loughborough University, UK. A.M.K. is grateful to the
Royal Society for financial support.

\end{document}